%% file: paper.tex
\documentclass[manuscript]{aastex}
\usepackage[utf8]{inputenc}
\usepackage{graphicx}
\usepackage{amsmath} % split in equation enviroment 

\newcommand{\X}{\textsf{X}}
\newcommand{\zdot}{\makebox[0pt][l]{.}} % arcmin with decimal dot
\newcommand{\up}[1]{\ifmmode^{\rm #1}\else$^{\rm #1}$\fi}
\newcommand{\uph}{\up{h}}
\newcommand{\upm}{\up{m}}
\newcommand{\arcd}{\ifmmode^{\circ}\else$^{\circ}$\fi}
\newcommand{\arcm}{\ifmmode{'}\else$'$\fi}
\newcommand{\arcs}{\ifmmode{''}\else$''$\fi}

\slugcomment{\today{} version}

\shorttitle{Proper motions in \X structure}
\shortauthors{Poleski et al.}

\begin{document}


\title{Asymmetric streaming motion in the Galactic bulge \X-shaped structure revealed by the OGLE-III proper motions}

\author{Rados\l{}aw Poleski\altaffilmark{1,2}, Andrzej Udalski\altaffilmark{2}, Andy Gould\altaffilmark{1},}
\author{M. K. Szyma\'nski\altaffilmark{2}, I. Soszy\'nski\altaffilmark{2}, M. Kubiak\altaffilmark{2}, G. Pietrzyński\altaffilmark{2,3}} 
\author{K. Ulaczyk\altaffilmark{2} and \L{}. Wyrzykowski\altaffilmark{2,4}}
\email{poleski@astronomy.ohio-state.edu}

\altaffiltext{1}{Department of Astronomy, Ohio State University, 140 W. 18th Ave., Columbus, OH 43210, USA}
\altaffiltext{2}{Warsaw University Observatory, Al. Ujazdowskie 4, 00-478 Warszawa, Poland}
\altaffiltext{3}{Universidad de Concepci\'on, Departamento de Astronomia, Casilla 160–C, Concepci\'on, Chile}
\altaffiltext{4}{Institute of Astronomy, University of Cambridge, CB3 0HA, UK}


\begin{abstract}
The Galactic bulge shows a double red clump in sight lines with $|b| \gtrsim 5^{\circ}$ and $-3^{\circ} \lesssim l \lesssim 4^{\circ}$. 
It is interpreted as a signature of an \X-shaped structure seen almost edge-on. 
We measure the proper motions of the stars belonging to the closer and the further arm of the \X-shaped structure. 
The intrinsic kinematic properties of the two arms are found by incorporating information taken from the luminosity function. 
At $b = -5^{\circ}$ we find that the proper motion difference between the two arms is a linear function of the Galactic longitude for $-0.1^{\circ} < l < 0.5^{\circ}$, which 
we interpret as a streaming motion of the stars within the \X-shaped structure. 
Such a streaming motion has not previously been reported. 
The proper motion difference is constant for $-0.8^{\circ} < l < -0.1^{\circ}$,
which gives us an estimate of bulge rotation speed of 
$87.9\pm8.2~{\rm km~s^{-1}~kpc^{-1}}$.

\end{abstract}



\section{Introduction} % ########################################################################

The bulge of the Galaxy is the closest such structure and the only one 
in which stars can be well resolved and studied individually. 
This in turn allows detailed investigation of properties of stellar populations within the bulge. 
The Galactic bulge was not formed via major mergers but was developed via bar buckling \citep{shen10}. 
It was recently found \citep{nataf10,mcwilliam10} that the Galactic bulge contains an \X-shaped structure.
This structure is aligned with a bar, which makes $\approx 30^{\circ}$ angle with Sun -- Galactic center direction, 
and its arms are significantly extended along the line of sight direction. 
The color-magnitude diagrams (CMDs) constructed in some bulge areas show a double-peaked red clump (RC) because the line of sight crosses two arms of the \X-shaped structure. 
In fields close to the Galactic disk the two RCs merge and appear single-peaked. 
These include the well studied low-extinction Baade window at Galactic coordinates $(l,b) = (1^{\circ},-3.9^{\circ})$.
The RC is also single-peaked in the areas where only one arm is seen. 
\X-shaped structures are known in other galaxies \citep{whitmore88} and may be centered (four arms cross at the center of the galaxy)
or off-centered \citep[arms cross in two areas within the disk that do not coincide with the Galactic center;][]{bureau06}. 

The double RC in the Galactic bulge was confirmed using 2MASS and VVV near infrared data by \citet{saito11,saito12}. 
More recently \citet{cao13} analyzed the spatial density of the RC stars and found a clear signature of the \X-shaped structure.
\citet{li12} found properties of the \X-shaped structure in the bulge model of \citet{shen10} to be consistent with observations. 

Until now there have been relatively few studies devoted to the kinematics of both RCs. 
In some cases they were conducted at fields where the \X-shaped structure is present but hard to be distinguished i.e., $-5^{\circ} \lesssim b \lesssim -3^{\circ}$. 
Three such examples 
were reported by \citet{mcwilliam10}.
The catalog of proper motions by \citet{sumi04} based on the second phase of 
the Optical Gravitational Lensing Experiment (OGLE-II) data was used 
to calculate a proper motion difference of $1.0\pm0.06$ milliarcseconds per year (${\rm mas~yr^{-1}}$)
between the two arms of the \X-shaped structure at the Baade window. 
The catalog of proper motions in the Plaut field $(l,b) = (0^{\circ},-8^{\circ})$ compiled by \citet{vieira07} resulted in a proper motion difference 
in longitude of $0.51\pm0.18~{\rm mas~yr^{-1}}$,
and of $0.19\pm0.19~{\rm mas~yr^{-1}}$ in latitude.
\citet{mcwilliam10} claimed also a $3.4\sigma$ difference in the latitudinal proper motion dispersion without providing 
the figures themselves. 
These authors also recalled the radial velocity (RV) differences between the faint and bright part of the RC obtained by \citet{rangwala09}: 
$-40\pm11~{\rm km~s^{-1}}$ at $(l,b) = ( 5.5^{\circ},-3.5^{\circ})$,
%$-4 \pm15~{\rm km~s}$ at $(l,b) = ( 1.1^{\circ},-3.9^{\circ})$ and
$-4 \pm15~{\rm km~s^{-1}}$ at the Baade window, and
$-32\pm11~{\rm km~s^{-1}}$ at $(l,b) = (-5.0^{\circ},-3.5^{\circ})$.
As pointed out by \citet{mcwilliam10} the results in the two fields with $l\approx \pm 5^{\circ}$ have magnitudes that are very close to the estimates of the model predictions by \citet{mao02} but have the opposite sign. 
The Baade window was also studied by \citet{babusiaux10}. 
An RV difference of $70\pm30~{\rm km~s^{-1}}$ was found after excluding metal-poor stars. 
The proper motion differences found by \citet{babusiaux10} in the Baade window based on the \citet{sumi04} catalog were not statistically significant. 

\citet{depropis11} measured a nonsignificant RV difference of $12\pm10~{\rm km~s^{-1}}$ between two RCs at the Plaut window.
\citet{ness12} measured the RV offset of $-30\pm12~{\rm km~s^{-1}}$ at $(l,b) = (0^{\circ},-5^{\circ})$ and $7\pm9~{\rm km~s^{-1}}$ in joint data for two fields: $(l,b) = (0^{\circ},-7.5^{\circ})$ and $(l,b) = (0^{\circ},-10^{\circ})$. 
Both figures were consistent with model predictions.
\citet{uttenthaler12} found $4.4\pm9.5~{\rm km~s^{-1}}$ RV difference at $(l,b) = (0^{\circ},-10^{\circ})$. 
The result changed to $5.2\pm9.5~{\rm km~s^{-1}}$ when the authors accounted for probabilities that each given star belongs to either RC. 
The small change of the result is surprising to us (see Sec.~\ref{sec:analysis}). 
Neither of the above mentioned studies searched for the changes in relative kinematics of the two arms on the scales smaller than $5^{\circ}$.

There is one problem inseparably involved in the kinematic studies of the Galactic bulge that use the RC stars. 
Not all the stars of the RC region of the CMD belong to the RC because there are underlying red giants 
that may be at different distances and thus posses different mean kinematics.
As noted by \citet{mao02} these stars dilute the measured kinematic differences between the brighter and fainter RC stars. 
There are two ways one can cope with this fact when comparing the observations with predictions based on Galactic models.
The first is to report the raw measured properties and compare them with model predictions, which are diluted in the same manner as observations \citep[see e.g.,\,][]{ness12}.
The disadvantage of this method is that the above mentioned results cannot be compared in detail to the models 
as long as one does not specify from which parts of the CMD stars were used in the calculations. 
The other approach is presented in this paper. 
The probability that a given star belongs to the brighter or fainter arm of the \X-shaped structure is assigned to each star 
(based on its extinction-corrected brightness) 
and using this we find the intrinsic properties of the two arms. 
The result can be directly compared to the model predictions, 
which are based upon the proper motions of stars selected based on their distances. 
It also allows  the analytical derivation of the Galactic properties.

We note one more proper motion study in bulge field, which does not show a double RC structure.
Based on the Hubble Space Telescope data \citet{clarkson08} found proper motions in a field $(l,b) = (1.25^{\circ},-2.65^{\circ})$. 
These authors assigned photometric distances to every star and calculated mean proper motions in distance bins. 
They concluded that the mean circular speed of the bulge stars follows solid body rotation to a cutoff radius of $0.3-0.4~{\rm kpc}$ with a maximum velocity of $25~{\rm km~s^{-1}}$. 
This corresponds to an angular rotation velocity of $62-83~{\rm km~s^{-1}kpc^{-1}}$. 
\citet{clarkson08} also estimated that biases in the stellar properties decrease the measured rotation velocity by a factor of two.
Thus the angular rotation velocity from that study is between $125$ and $167~{\rm km~s^{-1}kpc^{-1}}$.

The aim of this paper is to study the proper motions of the stars belonging to the two arms of the \X-shaped structure.
We use data collected by the OGLE-III survey to calculate the proper motions. 
The third phase of the OGLE project used a camera with better resolution than the second phase. 
In addition the observed sky-area was significantly larger and covered some fields with the double RC. 
The results presented here could not be achieved using the OGLE-II data. 
We describe the observations used in this study in Section~\ref{sec:obs}. 
The two following sections present the luminosity function construction and the calculation of proper motions of individual stars. 
Both of these are used in Sec.~\ref{sec:analysis} to find the intrinsic proper motion differences of stars in the two arms of the \X-shaped structure. 
We end by discussing the implications of our results.

\section{Observations} % ########################################################################
\label{sec:obs}

The OGLE-III project \citep{udalski03} was conducted with the 1.3-m Warsaw Telescope located at the Las Campanas Observatory, Chile. 
The telescope was equipped with an eight CCD chip mosaic camera. 
The total camera dimension was $8~{\rm k} \times 8~{\rm k}$ pixels, with a field of view $35\arcm \times 35\arcm$. 
The $15\mu m$ pixels gave a $0.26''$ pixel scale, allowing full benefit from the excellent seeing conditions. 
The observations were performed in $V$ and $I$ filters 
with 
the  majority of observations taken in the $I$ band. 
The $V$ band observations are used here only for color information. 
The exposure time was $120~{\rm s}$.

The main goal of the OGLE-III project were studies of the microlensing events which are effectively found only in the Galactic bulge. 
The observed bulge area was $92~{\rm square~degrees}$ within the range $-12^{\circ} < l < 13^{\circ}$ and $-7^{\circ} < b < 6^{\circ}$. 
During the survey observations, the sky area monitored decreased with time in order to increase the cadence in the fields showing the highest number of microlensing events. 
The number of epochs per field ranged from a dozen or so to $\approx 2400$, and the time span of observations ranged from about one year to eight years.
The photometric maps of the OGLE-III bulge fields were presented by \citet{szymanski11}. 
We select the fields where the double RC is well pronounced and observing coverage was good enough to obtain the precision of the proper motions below $1~{\rm mas~yr^{-1}}$ in each direction. 
We were left with three fields named BLG134, BLG167 and BLG176, 
the latter two of which are neighboring. 
The sky coordinates of the fields centers, the number of epochs collected and observing time span are summarized in Table \ref{tab:fields}.\
The sky area corresponding to the one CCD chip is called a subfield. 
Different subfields of a given field are distinguished by the number from 1 to 8 separated by a decimal point.

\begin{table}
\begin{center}
\caption{Characteristics of the observed fields.\newline \label{tab:fields}}
\begin{tabular}{lrrrrrr}
\tableline\tableline
\multicolumn{1}{c}{field name} &
\multicolumn{1}{c}{R.A.} &
\multicolumn{1}{c}{Dec.} &
\multicolumn{1}{c}{$l$} &
\multicolumn{1}{c}{$b$} &
\multicolumn{1}{c}{$N_{epoch}$} &
\multicolumn{1}{c}{$\Delta t$} \\
 &
\multicolumn{1}{c}{J2000.0} &
\multicolumn{1}{c}{J2000.0} &
\multicolumn{1}{c}{$[^{\circ}]$} &
\multicolumn{1}{c}{$[^{\circ}]$} &
 &
\multicolumn{1}{c}{$[{\rm yrs}]$} \\
\tableline
%BLG134 & 17\uph57\upm38\zdot\ups2 & $-34\arcd12\arcm14\arcs$ & $ -3.2362 $ & $ -4.8829 $ & 326 & 4.6 \\ 
%BLG167 & 18\uph03\upm32\zdot\ups6 & $-31\arcd50\arcm15\arcs$ & $ -0.5573 $ & $ -4.8001 $ & 360 & 4.6 \\
%BLG176 & 18\uph06\upm08\zdot\ups9 & $-31\arcd14\arcm55\arcs$ & $  0.2313 $ & $ -4.9995 $ & 355 & 4.5 \\
BLG134 & 17\uph57\zdot\upm6 & $-34\arcd12\arcm$ & $ -3.24 $ & $ -4.88 $ & 326 & 4.6 \\ 
BLG167 & 18\uph03\zdot\upm5 & $-31\arcd50\arcm$ & $ -0.56 $ & $ -4.80 $ & 360 & 4.6 \\
BLG176 & 18\uph06\zdot\upm4 & $-31\arcd15\arcm$ & $  0.23 $ & $ -5.00 $ & 355 & 4.5 \\
\tableline
\end{tabular}
%% Any table notes must follow the \end{tabular} command.
%\tablecomments{We can also attach a long-ish paragraph of explanatory material to a table.}
\end{center}
\end{table}

The astrometric measurements analyzed here are independent of the standard OGLE-III reduction, which used Difference Image Analysis. 
In order to measure the centroids of the stars, we implemented the effective PSF approach presented by \citet{anderson00} and \citet{anderson06}. 
The effective PSF is a convolution of the instrumental profile with the pixel sensitivity map. 
Both of the latter are unknown, but they are not needed as the stellar centroids can be found using an accurate effective PSF. 
In our implementation the grid on which the effective PSF is defined supersamples the pixel by a factor of four in each direction.
The spatial variation of the profile is treated by calculating it on a $2\times4$ grid for each CCD chip individually. 
The effective PSF for a given place of the CCD chip is evaluated by a linear interpolation of the four nearest grid profiles.

\section{Luminosity function} % ########################################################################
\label{sec:lf}

\begin{figure}
\plotone{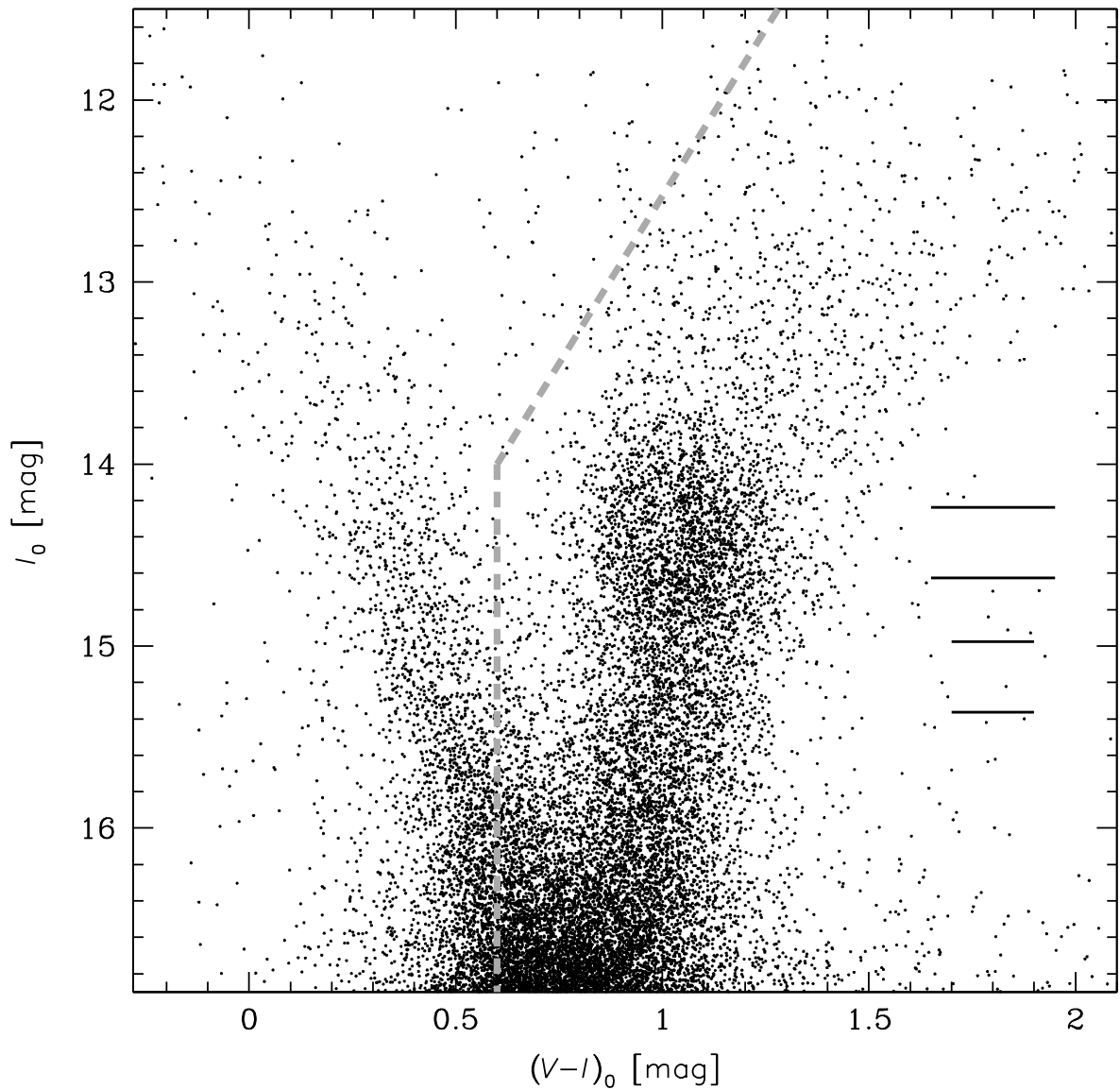}
\caption{Example of a derredened CMD (for the subfield BLG175.6). 
Stars lying to the right of the gray dashed line are used in construction of the luminosity function.
Two longer horizontal lines mark the brightnesses of the two RCs. 
The two shorter horizontal lines mark the brightnesses of the two RGBBs.
\label{fig:cmd}}
\end{figure}

In order to assign each given star a probability that it belongs to the brighter or fainter arm we deredden each star individually using 
the interpolated extinction maps of \citet{nataf12tmp}. 
The $E(V-I)$ reddening varies between $0.53~{\rm mag}$ and $0.81~{\rm mag}$ for BLG134 and between $0.62~{\rm mag}$ and $1.29~{\rm mag}$ for the other two fields. 
Figure~\ref{fig:cmd} presents a sample CMD after the extinction correction was applied 
($I_0$ and $(V-I)_0$ are extinction corrected values of $I$ band brightness and $(V-I)$ color). 
We use a two-dimensional extinction map,
so the foreground disk stars are artificially shifted toward bluer colors and brighter magnitudes. 

\begin{figure}
\epsscale{.72}
\plotone{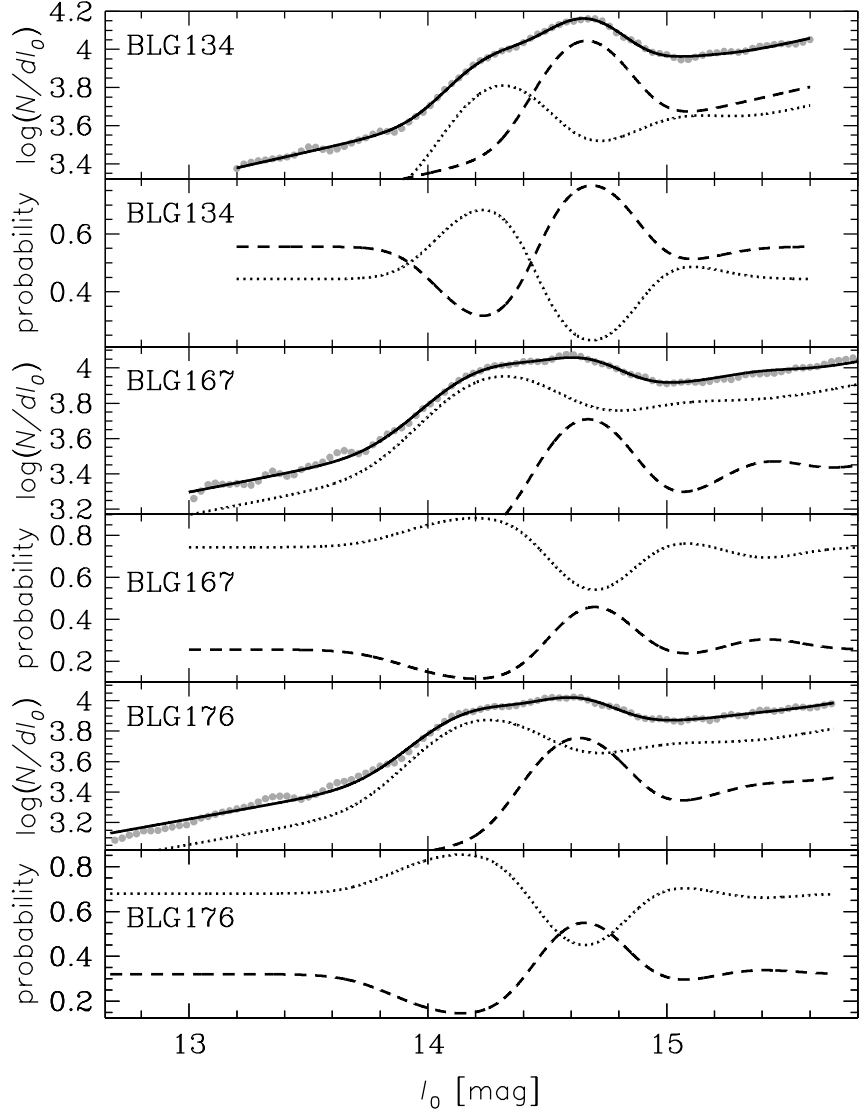} 
\caption{Luminosity functions and derived probabilities as a function of extinction-corrected $I_0$ brightness.
The odd panels (counting from top) present the luminosity functions (gray points) and corresponding fits (solid line). 
The dotted and dashed lines present the contribution of the brighter and fainter arm, respectively. 
The even panels show the probability that the star of given brightness belongs to the brighter (dotted line) or fainter(dashed line) arm of the \X-shaped structure. 
\label{fig:lf}}
\end{figure}

The luminosity function of the bulge red giants is constructed by taking into account the stars that are redder than the gray line in Fig. \ref{fig:cmd}. 
This line changes its slope at $I_0 = 14~{\rm mag}$ to exclude the disk RC stars, 
which are abundant in the CMD region $(V-I)_0 \approx 0.8~{\rm mag}$ and $I_0 < 13.5~{\rm mag}$. 
We argue that this selection of stars removes the great majority of the foreground disk red giants. 
The nearby red giants are shifted toward bluer colors by extinction correction and thus are not included in our selection box. 
We verified this using the Besan{\c c}on Galactic model \citep{robin03}. 
Additional cleaning of the sample is conducted later on, when the objects with the highest proper motions are removed. 
A separate luminosity function was constructed for each field. 
All are shown in three panels of Fig. \ref{fig:lf}.
We found that calculation of a separate luminosity function of each OGLE field is close to the optimum between obtaining good statistics and
characterizing the spatial changes in the luminosity function. 

The luminosity function for each field is fitted with an analytical function 
that takes into account both arms of the \X-shaped structure. 
In previous studies \citep[e.g.,\,][]{nataf11} simpler functions were used. 
The number of stars in each arm is a sum of three components: 
red giant branch (represented by an exponential function), 
RC (represented by a Gaussian) 
and the red giant branch bump \citep[RGBB henceforth;][; represented by a Gaussian]{gallart98,nataf11}. 
We take into account the RGBB and ignore the asymptotic giant branch bump which was also found in the Galactic bulge \citep{nataf11} because of two major differences between the two structures. 
First, the number counts of the asymptotic giant branch bump are at least an order of magnitude smaller.
Second, the absolute brightness difference between the RGBB and the RC is significantly smaller than for the asymptotic giant branch bump ($0.737~{\rm mag}$ compared to $1.06~{\rm mag}$). 
The function that we fit has sixteen parameters (two for each of the exponentials and three for each Gaussian), but we make several assumptions to reduce the number of fitted parameters.
The slopes of the exponential components were kept the same. 
The number of the red giants at brightness of the RC was also kept the same. 
The dispersions of the Gaussians representing the RGBB were the same as the dispersions of the Gaussian representing RC in each of the arms. 
The brightness difference between RGBB and RC as well as the number ratio of RGBB to RC stars were fixed at values found by 
\citet{nataf13} of $0.737~{\rm mag}$ and $0.201$, respectively. 
These parameters depend on the metallicity and thus their true values in the analyzed fields may be different than assumed. 
We made two exceptions to the fitting procedure described above, which significantly improved our fits.
First, in the field BLG134 the contribution of the fainter RGBB was neglected.
Second, in the field BLG176 the number of the RGBB stars relative to the RC stars in the fainter arm was a free parameter with the best fitting value of $0.075\pm0.063$, compared to $0.201$ in the standard fit.
The fits had eight or nine free parameters and between 81 and 101 % blg134 - 81, blg167 - 93, blg176 - 101
data points in the luminosity function.
We adjust the interval of $I_0$ in which the fit is performed for each field separately. 
The difference in the extinction-corrected brightness of the two RCs are 
$0.353\pm0.034~{\rm mag}$, %14.657536 0.015362 14.304495 0.030436
$0.365\pm0.051~{\rm mag}$, and %14.665016 0.027791 14.299995 0.041915
$0.387\pm0.046~{\rm mag}$ % 14.625905 0.025978 14.238416 0.037846
for BLG134, BLG167, and BLG176, respectively.

Each star was assigned a probability that it belongs to the brighter ($p_b(I_0)$) or the fainter ($p_f(I_0)$) arm of the \X-shaped structure. 
This probability is equal to the ratio of number of stars from a given arm to the total number of stars, both of which are the values of the fitted function at the extinction-corrected brightness of the star. 
The plots of $p_b(I_0)$ are shown on the three panels of Fig. \ref{fig:lf}.

\section{Calculation of proper motions} % ########################################################################
\label{sec:pm}

The crucial point in the calculation of  high accuracy proper motions is finding an adequate transformation of grids. 
We note that the OGLE-III observing strategy was not optimized for precise astrometry: 
there was typically only one observation of a given field taken per night,  
the seeing FWHM varied up to a factor of 2.5, 
exposures were taken at different airmasses, and the airmass could change significantly between any two consecutive exposures. 
These issues prevent us from using a detailed geometric correction common for all of the subfields of the given exposure as was done by \citet[e.g., ][]{bellini10}.
Instead, we derived the transformation grids for each subfield individually. 
This enforced the proper motion zero points to be different for each subfield. 

The cross-matching of the star catalogs from each image was done using a simplified version of the algorithm presented by \citet{pal06}. 
Second order polynomial grids were calculated in order to transform the measured positions to a common frame. 
The stellar centroids from all the frames of a given subfield were transformed to this common frame.
After that, we fit a model for each star that took into account the coordinates for a selected epoch, the proper motion and the differential refraction coefficient. 
The value of differential refraction shift is calculated by multiplying the coefficient by the tangent of the zenith distance. 
The zero point of the proper motions is set by calculating the mean of proper motions for red giants brighter than $I_0 = 13.8~{\rm mag}$. 
Stars with proper motions significantly higher than those of bulge stars are not considered. 
From the proper motion of every star the mean proper motion of bulge stars is subtracted.
These proper motions and differential refraction coefficients 
are used to calculate the residuals of the stellar positions. 
Those are transformed back from the common frame to the individual frames and the measured positions are subsequently corrected for these residuals. 
Then, new grid transformations are found and using them we calculate the mean position of every reference star.
If these differ more than $0.15~{\rm pix}$ from the initial common frame, we remove the star from the list of reference objects. 
The mean positions are used as a new common frame, which in principle is corrected for differential refraction. 
The grid transformations from the individual frames to that frame are found and final fits of positions for a selected epoch, proper motions and differential refraction coefficients are performed. 

The uncertainties of the proper motions are found using the bootstrap method \citep{press92}. 
From a set of exposures of a given field, we draw with replacement a subset whose number of elements is equal to the original set. 
Multiple such subsets are drawn and for each of them the procedure of calculating the proper motions described above is repeated.
The standard deviation of the proper motion of a given star is taken as the measurement uncertainty. 
The proper motion uncertainties are about $0.3~{\rm mas~yr^{-1}}$ for $I = 14~{\rm mag}$ and about $0.5~{\rm mas~yr^{-1}}$ for $I = 16~{\rm mag}$. 
It was found that the proper motion uncertainties increase near the edges of subfields. 
This is caused by the less well defined grid transformations in these parts of the subfields. 
We compared the proper motions measured for stars present in overlapping parts of the adjacent subfields and found that bootstrap estimates are consistent with measured proper motion differences. 
The raw measurements of proper motions used in this study will be published as a part of the proper motion catalog covering the whole OGLE-III bulge area.

\section{Analysis} % ########################################################################
\label{sec:analysis}

The intrinsic (i.e., undiluted by red giants) proper motion difference between the brighter and fainter RC can be found using the calculated proper motions and probabilities that a given star belongs to either the brighter or fainter arm.
Let index $i$ label the stars. 
For each of them we have measured the proper motion in Galactic coordinates: $\mu_{i,l\star}$, $\mu_{i,b}$\footnote{The $\mu_{l\star}=\mu_l\cos b$ and $\mu_b$ are the proper motions in Galactic coordinate system with longitude value corrected for scale changes.} 
and corresponding uncertainties $\xi_{i,l\star}$, $\xi_{i,b}$. 
We know also the extinction-corrected brightness $I_{i,0}$ which gives the probabilities $p_b(I_{i,0})$ and $p_f(I_{i,0})$.
For the fainter arm the average (dispersion) of the proper motions in Galactic longitude is denoted $\mu_{{\rm f},l\star}$ ($\sigma_{{\rm f},l\star}$), while for Galactic latitude the corresponding value is $\mu_{{\rm f},b}$ ($\sigma_{{\rm f},b}$). 
For the brighter arm the corresponding symbols have first index changed from ${\rm f}$ to ${\rm b}$. 
The likelihood function for a single star ($\mathcal{L}_i$) is defined as:

\begin{equation}\begin{split}
\mathcal{L}_i = 
\frac{p_b(I_{i,0})}{2\pi\sqrt{\left(\sigma_{{\rm b},l\star}^2+\xi_{i,l\star}^2\right)\left(\sigma_{{\rm b},b}^2+\xi_{i,b}^2\right)}}\exp\left(-\frac{(\mu_{i,l\star}-\mu_{{\rm b}, l\star})^2}{2(\sigma_{{\rm b},l\star}^2+\xi_{i,l\star}^2)}-\frac{(\mu_{i,b}-\mu_{{\rm b}, b})^2}{2(\sigma_{{\rm b},b}^2+\xi_{i,b}^2)}\right) + \\
\frac{p_f(I_{i,0})}{2\pi\sqrt{\left(\sigma_{{\rm f},l\star}^2+\xi_{i,l\star}^2\right)\left(\sigma_{{\rm f},b}^2+\xi_{i,b}^2\right)}}\exp\left(-\frac{(\mu_{i,l\star}-\mu_{{\rm f}, l\star})^2}{2(\sigma_{{\rm f},l\star}^2+\xi_{i,l\star}^2)}-\frac{(\mu_{i,b}-\mu_{{\rm f}, b})^2}{2(\sigma_{{\rm f},b}^2+\xi_{i,b}^2)}\right)
\end{split}\end{equation}

The product of likelihoods for all the stars in a given subfield is the function we maximize using the Markov Chain Monte Carlo (MCMC). 
A separate chain was run for each subfield. 
After trial and error we choose the same interval of $I_0$ brightness between $14~{\rm mag}$ and $15~{\rm mag}$.
In this range both RCs are prominent, 
and that is where the most information on the undiluted proper motions comes from. 
For the stars brighter than $I_0 = 14~{\rm mag}$ the probabilities are poorly constrained, and the number of brighter stars is smaller. 
For the stars fainter  than $I_0 = 15~{\rm mag}$, the RGBBs significantly contribute to the luminosity function and the analytical fits presented in Sec.~\ref{sec:lf} are slightly poorer. 
For $I_0\approx 15.7~{\rm mag}$, the main sequence disk stars start to contribute significantly to the luminosity function. 
The mean value of the effective number of stars in the brighter (fainter) arm i.e., $\sum_i p_b(I_{i,0})$ ($\sum_i p_f(I_{i,0})$) per subfield is 1093 (1499) in the field BLG134.
In the field BLG167 the corresponding value is 1723 (669), and for BLG176 it is 1416 (771).
The values in individual subfields do not differ by more than $10\%$ except the subfields BLG167.5 ($15\%$ larger than the mean) and BLG176.5 ($13\%$ larger than the mean).

\begin{figure}
\plotone{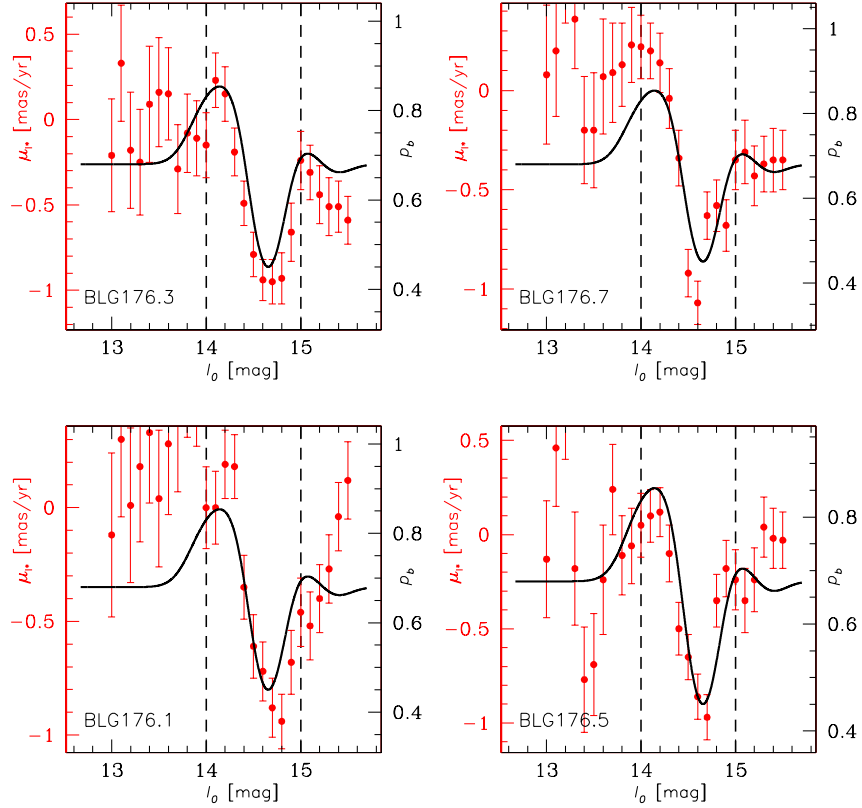} 
\caption{Comparison of $\mu_{l\star}$ 
(binned in overlapping $0.2~{\rm mag}$ wide bins which are $0.1~{\rm mag}$ apart) 
 and probabilities derived as a function of the dereddened brightness. 
Each panel presents a different subfield. 
The dashed vertical lines show the range $14~{\rm mag} < I_0 < 15~{\rm mag}$ that we use in the fit. 
\label{fig:pmu}}
\end{figure}

In order to illustrate the reliability of our MCMCs, we overlay the $p_b(I_0)$ and $\mu_{i,l\star}$ averaged in $0.1~{\rm mag}$ wide $I_0$ bins. 
The panels of Fig.~\ref{fig:pmu} present such plots for four sample subfields. 
The agreement between the measured proper motions and the $p_b$ for $I_0$ between $14~{\rm mag}$ and $15~{\rm mag}$ proves the consistency of our procedures of extinction correction, luminosity function construction, fitting analytical model to it, and proper motion calculation.

The results of the MCMC are presented in Table~\ref{tab:res}. 
We do not state values of $\mu_{{\rm b},l\star}$, $\mu_{{\rm b},b}$, $\mu_{{\rm f},l\star}$, and $\mu_{{\rm f},b}$ as their zero points may change in different subfields. 
Instead we present their differences i.e.,\, $\Delta\mu_{l\star} = \mu_{{\rm b},l\star} - \mu_{{\rm f},l\star}$ and 
$\Delta\mu_{b} = \mu_{{\rm b},b} - \mu_{{\rm f},b}$, which carry all the astrophysical information.

We note that the differences of $\mu_{l\star}$ at the brightness of the two RCs are up to $1.2~{\rm mas~yr^{-1}}$ in the subfields presented in Fig.~\ref{fig:pmu}. 
The values of $\Delta\mu_{l\star}$, which are corrected for dilution by red giants, for these subfields are about twice larger. 
Such a large ratio is caused by the fact that even for the brightness of the brighter RC some contribution of the fainer arm is seen and vice versa. 
Among the previous comparisons of two arms kinematics, only \citet{uttenthaler12} accounted for the probabilities that  measured stars belong to the either RC. 
In contrast to us they found a very small change of the kinematic properties.

\clearpage

\begin{deluxetable}{lrrrrrrrr}
\tabletypesize{\scriptsize}
\tablecaption{Proper motion statistics of two arm of the \X-shaped structure\label{tab:res}}
\tablewidth{0pt}
\tablehead{
\colhead{subfield} & 
\colhead{$l [^{\circ}]$} & \colhead{$b[^{\circ}]$} & 
\colhead{$\Delta\mu_{l\star}$} & \colhead{$\sigma_{{\rm b},l\star}$} & \colhead{$\sigma_{{\rm f},l\star}$} & 
\colhead{$\Delta\mu_{b}$} & \colhead{$\sigma_{{\rm b},b}$} & \colhead{$\sigma_{{\rm f},b}$}  
}
\startdata
\input{res_mu_tab}
\enddata
%% Text for table notes should follow after the \enddata but before
%% the \end{deluxetable}. Make sure there is at least one \tablenotemark
%% in the table for each \tablenotetext.
\tablecomments{First three columns give the name of the subfield and the Galactic coordinates of its center. All proper motion differences and disperions are in ${\rm mas~yr}^{-1}$. 
} 
%\tablenotetext{a}{Sample footnote for table~\ref{tbl-1} that was generated with the deluxetable environment}
%\tablenotetext{b}{Another sample footnote for table~\ref{tbl-1}}
\end{deluxetable}

All analyzed fields are close to $b = -5^{\circ}$.
The range of Galactic longitudes is significantly larger. 
The fields BLG167 and BLG176 span from $l=-0.9^{\circ}$ to $l=0.6^{\circ}$. 
The field BLG134 is located around $2^{\circ}$ from these two and is therefore discussed separately.

\begin{figure}
\plotone{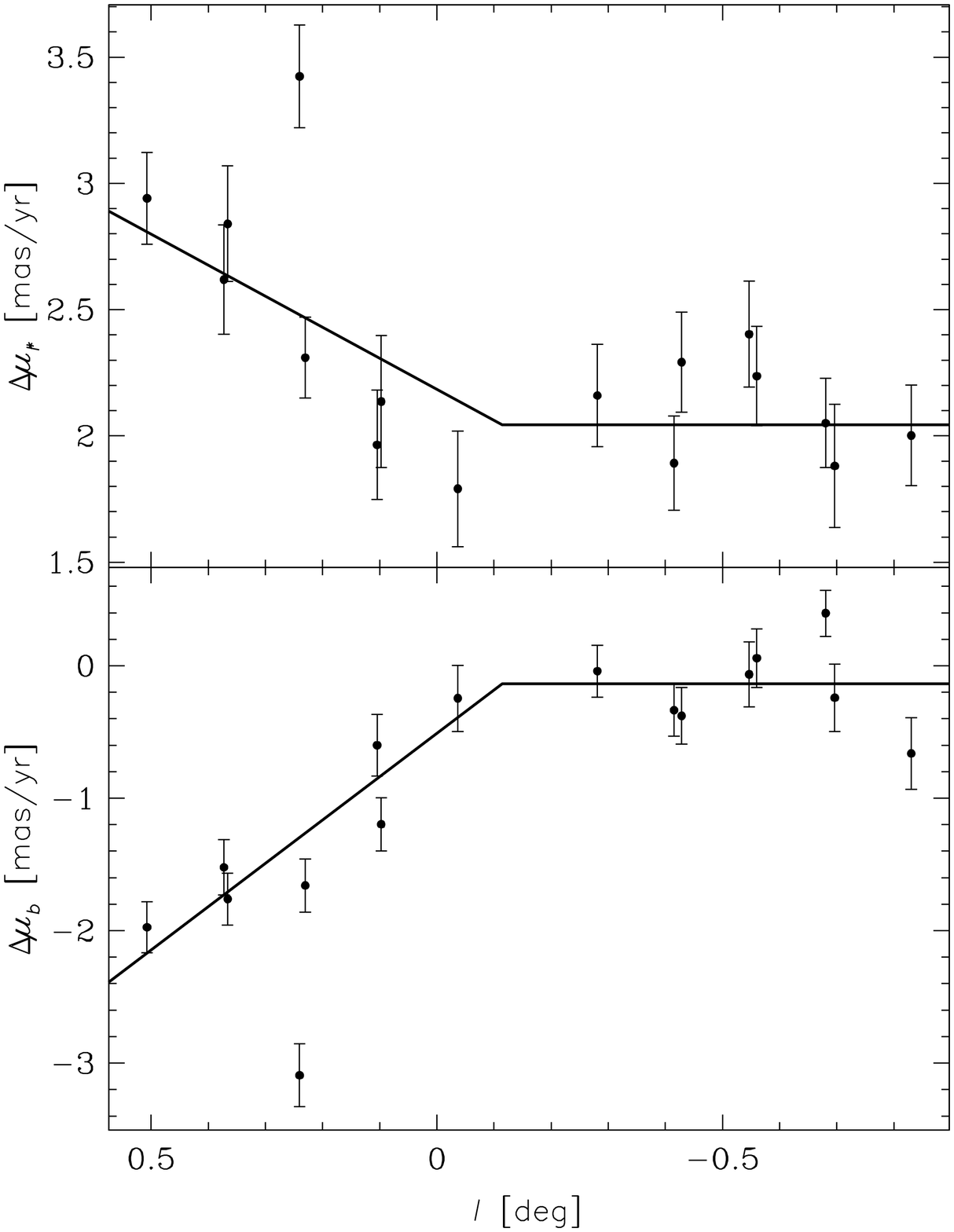} 
\caption{Proper motion differences of two arms of the \X-shaped structure as a function of the Galactic longitude. 
The upper panel shows the longitudinal proper motion difference while the lower one presents the latitudinal proper motion difference. 
The lines are fits with common longitude of the break point in both panels.
\label{fig:pmdiff}}
\end{figure}

Figure~\ref{fig:pmdiff} presents the proper motion differences $\Delta\mu_{l\star}$ and $\Delta\mu_{b}$ vs.~Galactic longitude. 
We are analyzing the fields close to $l=0^{\circ}$, so the purely cylindrical rotation identified from RV surveys \citep{kunder12,ness13} should result in constant values of $\Delta\mu_{l\star}$ and in $\Delta\mu_b$ being close to zero. 
However, this expectation actually only holds for negative $l$. 
For positive $l$, both $\Delta\mu_{l\star}$ and $\Delta\mu_{b}$ are linear functions of $l$.
We fit a five parameter phenomenological 
model to both $\Delta\mu_{l\star}(l)$ and $\Delta\mu_b(l)$. 
The subfield BLG176.2 is not included in this fit since $\Delta\mu_{l\star}$ and $\Delta\mu_b$ derived in this field are the most outlying points on both panels of Fig.~\ref{fig:pmdiff}.
For $l < l_{\rm break}$ we assume

$$\Delta\mu_{l\star}(l) = \Delta\mu_{l\star}^{const}$$
$$\Delta\mu_{b}(l) = \Delta\mu_{b}^{const}$$
and for $l \geqslant l_{\rm break}$ we assume
$$\Delta\mu_{l\star}(l) = \Delta\mu_{l\star}^{const} + \frac{d\,\Delta\mu_{l\star}}{d\,l}\cdot\left(l-l_{\rm break}\right)$$
$$\Delta\mu_{b}(l) = \Delta\mu_{b}^{const} + \frac{d\,\Delta\mu_{b}}{d\,l}\cdot\left(l-l_{\rm break}\right).$$

The fit resulted in $l_{\rm break}$ value of $-0.114^{\circ}\pm0.060^{\circ}$, i.e.,\,$2\sigma$ away from $0^{\circ}$. 
The parameters for the  longitudinal proper motion are $\Delta\mu_{l\star}^{const} = 2.04\pm0.07~{\rm mas~yr^{-1}}$ and $\frac{d\,\Delta\mu_{l\star}}{d\,l} = 1.23\pm0.32~{\rm mas~yr^{-1}~deg^{-1}}$. 
For the latitudinal proper motion we obtain $\Delta\mu_{b}^{const} = -0.14\pm0.08~{\rm mas~yr^{-1}}$ and $\frac{d\,\Delta\mu_{b}}{d\,l} = -3.27\pm0.43~{\rm mas~yr^{-1}~deg^{-1}}$. 
The fit resulted in $\chi^2/{\rm dof} = 42.1/25$. % l* contribution = 14.54; b contribution 27.57; (14.54 + 27.57)/(30-5) = 1.6844 [30 because blg176.2 is not taken into account]

\section{Discussion} % ########################################################################
\label{sec:discus}

The most striking result of Fig.~\ref{fig:pmdiff} is the presence of significant changes of the proper motion differences with the Galactic longitude for $l>-0.1^{\circ}$. 
We interpret this as a signature of asymmetric streaming motions of stars in the \X-shaped structure. 
The fact that the mean motion of the stars belonging to a certain structure is different from the mean motion of this structure is known e.g.,\, in the case of Galactic spiral arms. 
The measurements presented 
in this paper are insufficient to determine 
in which of the arms the streaming motion occurs. 
None of the published models of the \X-shaped structure gives predictions of the streaming motions that we report.

The value of $\Delta\mu^{const}_{l\star}$ can be used to constrain the angular velocity of the Galactic bar. 
For small values of distance modulus differences of two RCs $\Delta I_{\rm RC}$ their relative distances can be approximated by 
$\frac{{\rm ln}10\Delta I_{RC}}{-5}$. 
The bar angular velocity depends on the proper motion difference measured and the longitudinal proper motion of Sgr A*:
$$\Omega~=~\mu_{{\rm Sgr\,A*}, l\star}+\frac{5\Delta\mu_{l\star}^{\rm const}}{{\rm ln}10\Delta I_{RC}}$$
Substituting the $\Delta\mu_{l\star}^{\rm const}$, $\Delta I_{RC}$ in BLG167 of $-0.365\pm0.051~{\rm mag}$ and 
%$\mu_{{\rm Sgr~A*}, l\star}=-6.379\pm0.026~{\rm mas~yr^{-1}}$
$\mu_{{\rm Sgr~A*}, l\star}=-6.38~{\rm mas~yr^{-1}}$
as measured by \citet{reid04} we obtain 
$\Omega = -18.5\pm1.8~{\rm mas~yr^{-1}} = -87.9\pm8.2~{\rm km~s^{-1}~kpc^{-1}}$\footnote{Negative sign indicates the same direction of bulge rotation as the Sun's orbital motion.}.
This result is marginally consistent with results of \citet{clarkson08} without correcting for biases in their analysis.
If we assume that the Galactic angular velocity is constant up to some Galactic radius $R'$ and for larger radii the circular speed is the same as the local circular velocity of the Sun $V_c = 238\pm9~{\rm km~s^{-1}}$ \citep{schonrich12} then we can estimate $R' = V_c/\left|\Omega\right| = 2.71\pm0.28~{\rm kpc}$. % 2.7076 +/- 0.27255
The bar angle relative to Sun--Galactic center line is $\approx 30^{\circ}$ \citep[e.g.,\,][]{cao13}. 
Thus we can estimate that the bar points that  
are $2.71\pm0.28~{\rm kpc}$ away from the Galactic center are observed at $l=12.87\pm 2.4^{\circ}$ and $l=-7.27\pm0.44^{\circ}$
 (these uncertainties do not take into account the uncertainty of the bar angle).

The proper motion differences in the field BLG134 are similar to those found in field BLG167. 
The two most outlying subfields are BLG134.8 and BLG134.5. 
The former has the smallest longitude among the subfields analyzed here, while the latter is the closest to the Galactic disk among the  BLG134 subfields. 
Thus, these discrepancies are most likely caused by significantly smaller contribution of the brighter arm in these subfields.

The ratio of proper motion dispersions
in both arms along
the latitudinal direction can be used as an independent estimate of the distance ratio of the two arms if the same tangential velocity distribution is assumed in each arm. 
We note that this may not be the case since in a given sight line the two arms are at the different distances from the Galactic disk.

There are several studies devoted to the proper motion dispersions in the Galactic bulge \citep[][and references therein]{spaenhauer92,kuijken02,vieira07}. 
These dispersions were found to be anisotropic with $\sigma_{l\star}/\sigma_b \approx 1.2$. 
We do not confirm this finding when the two arms are treated separately. 
Analysis of the posterior probability distribution of the MCMC revealed that only in five cases did the ratios differ significantly from unity.
In each of these cases the longitudinal proper motion dipsersion is smaller than the latitudinal one. 
The value of the dispersion ratio in the brighter arm i.e.\, $\sigma_{{\rm b},l\star}/\sigma_{{\rm b},b}$ is $0.868\pm0.032$ in BLG176.8 and $0.774\pm0.025$ in BLG167.8, while for the fainter arm $\sigma_{{\rm f},l\star}/\sigma_{{\rm f},b}$ is
$0.763\pm0.071$, $0.707\pm0.072$, and $0.709\pm0.063$ for BLG176.8, BLG167.8, and BLG167.7, respectively.

\acknowledgments
We thank A.~Robin and S.~Koz\l{}owski for discussion. 
This work was supported by Polish Ministry of Science and Higher Education through the program ''Iuventus Plus'' award No. IP2011 043571 to RP. 
AG acknowledges supported by NSF grant AST 1103471.
The OGLE project has received funding from the European Research Council under the European Community’s Seventh Framework Programme (FP7/2007-2013)/ERC grant agreement No.~246678 to AU.

\bibliographystyle{apj}
%\bibliography{paper}

\end{document}

%% file: res_mu_tab.tex
BLG134.1 & $ -3.358 $ & $ -5.129 $ & $  1.95 \pm 0.23 $ & $ 2.90 \pm 0.10 $ & $ 1.91 \pm 0.07 $ & $ -0.11 \pm 0.19 $ & $ 2.68 \pm 0.09 $ & $ 1.85 \pm 0.08 $ \\
BLG134.2 & $ -3.226 $ & $ -5.053 $ & $  1.71 \pm 0.23 $ & $ 3.04 \pm 0.10 $ & $ 2.21 \pm 0.08 $ & $ -0.08 \pm 0.21 $ & $ 2.97 \pm 0.10 $ & $ 1.87 \pm 0.08 $ \\
BLG134.3 & $ -3.095 $ & $ -4.976 $ & $  1.72 \pm 0.25 $ & $ 2.97 \pm 0.11 $ & $ 1.99 \pm 0.08 $ & $  0.23 \pm 0.20 $ & $ 2.67 \pm 0.11 $ & $ 2.01 \pm 0.08 $ \\
BLG134.4 & $ -2.963 $ & $ -4.900 $ & $  1.81 \pm 0.19 $ & $ 2.87 \pm 0.09 $ & $ 1.88 \pm 0.07 $ & $ -0.18 \pm 0.18 $ & $ 2.78 \pm 0.08 $ & $ 1.78 \pm 0.06 $ \\
BLG134.5 & $ -3.108 $ & $ -4.646 $ & $  3.25 \pm 0.20 $ & $ 2.84 \pm 0.13 $ & $ 2.28 \pm 0.07 $ & $ -1.93 \pm 0.17 $ & $ 2.84 \pm 0.11 $ & $ 2.27 \pm 0.07 $ \\
BLG134.6 & $ -3.242 $ & $ -4.720 $ & $  2.25 \pm 0.24 $ & $ 2.82 \pm 0.12 $ & $ 1.84 \pm 0.07 $ & $ -0.14 \pm 0.21 $ & $ 2.78 \pm 0.11 $ & $ 2.13 \pm 0.09 $ \\
BLG134.7 & $ -3.375 $ & $ -4.795 $ & $  1.89 \pm 0.27 $ & $ 3.09 \pm 0.10 $ & $ 2.14 \pm 0.07 $ & $ -0.65 \pm 0.17 $ & $ 2.85 \pm 0.08 $ & $ 2.00 \pm 0.07 $ \\
BLG134.8 & $ -3.507 $ & $ -4.872 $ & $  3.37 \pm 0.18 $ & $ 2.38 \pm 0.10 $ & $ 2.31 \pm 0.08 $ & $ -2.74 \pm 0.19 $ & $ 2.26 \pm 0.10 $ & $ 2.62 \pm 0.09 $ \\
BLG167.1 & $ -0.680 $ & $ -5.039 $ & $  2.05 \pm 0.18 $ & $ 2.73 \pm 0.07 $ & $ 1.45 \pm 0.11 $ & $  0.40 \pm 0.18 $ & $ 2.63 \pm 0.07 $ & $ 1.76 \pm 0.12 $ \\
BLG167.2 & $ -0.547 $ & $ -4.967 $ & $  2.40 \pm 0.21 $ & $ 2.73 \pm 0.07 $ & $ 1.83 \pm 0.11 $ & $ -0.07 \pm 0.25 $ & $ 2.58 \pm 0.07 $ & $ 2.02 \pm 0.13 $ \\
BLG167.3 & $ -0.415 $ & $ -4.898 $ & $  1.89 \pm 0.19 $ & $ 2.79 \pm 0.07 $ & $ 1.46 \pm 0.10 $ & $ -0.34 \pm 0.20 $ & $ 2.54 \pm 0.06 $ & $ 1.83 \pm 0.13 $ \\
BLG167.4 & $ -0.281 $ & $ -4.821 $ & $  2.16 \pm 0.21 $ & $ 2.64 \pm 0.06 $ & $ 1.58 \pm 0.09 $ & $ -0.04 \pm 0.20 $ & $ 2.64 \pm 0.07 $ & $ 1.91 \pm 0.13 $ \\
BLG167.5 & $ -0.428 $ & $ -4.556 $ & $  2.29 \pm 0.20 $ & $ 2.82 \pm 0.07 $ & $ 1.78 \pm 0.11 $ & $ -0.38 \pm 0.22 $ & $ 2.76 \pm 0.07 $ & $ 2.01 \pm 0.14 $ \\
BLG167.6 & $ -0.560 $ & $ -4.633 $ & $  2.24 \pm 0.20 $ & $ 2.80 \pm 0.07 $ & $ 1.72 \pm 0.09 $ & $  0.06 \pm 0.23 $ & $ 2.87 \pm 0.06 $ & $ 2.03 \pm 0.14 $ \\
BLG167.7 & $ -0.696 $ & $ -4.706 $ & $  1.88 \pm 0.25 $ & $ 2.87 \pm 0.08 $ & $ 1.76 \pm 0.14 $ & $ -0.24 \pm 0.26 $ & $ 2.63 \pm 0.06 $ & $ 2.48 \pm 0.13 $ \\
BLG167.8 & $ -0.830 $ & $ -4.781 $ & $  2.00 \pm 0.20 $ & $ 2.49 \pm 0.07 $ & $ 1.75 \pm 0.13 $ & $ -0.66 \pm 0.28 $ & $ 3.22 \pm 0.08 $ & $ 2.49 \pm 0.17 $ \\
BLG176.1 & $  0.104 $ & $ -5.245 $ & $  1.97 \pm 0.22 $ & $ 2.84 \pm 0.08 $ & $ 1.80 \pm 0.10 $ & $ -0.60 \pm 0.24 $ & $ 2.79 \pm 0.07 $ & $ 1.94 \pm 0.12 $ \\
BLG176.2 & $  0.240 $ & $ -5.176 $ & $  3.42 \pm 0.21 $ & $ 3.33 \pm 0.10 $ & $ 1.97 \pm 0.10 $ & $ -3.09 \pm 0.24 $ & $ 3.20 \pm 0.10 $ & $ 2.23 \pm 0.12 $ \\
BLG176.3 & $  0.373 $ & $ -5.104 $ & $  2.62 \pm 0.22 $ & $ 2.69 \pm 0.10 $ & $ 1.90 \pm 0.09 $ & $ -1.52 \pm 0.21 $ & $ 2.61 \pm 0.08 $ & $ 1.84 \pm 0.10 $ \\
BLG176.4 & $  0.507 $ & $ -5.029 $ & $  2.94 \pm 0.19 $ & $ 2.62 \pm 0.08 $ & $ 2.02 \pm 0.09 $ & $ -1.97 \pm 0.20 $ & $ 2.61 \pm 0.08 $ & $ 2.30 \pm 0.10 $ \\
BLG176.5 & $  0.366 $ & $ -4.772 $ & $  2.84 \pm 0.23 $ & $ 2.77 \pm 0.09 $ & $ 2.21 \pm 0.10 $ & $ -1.76 \pm 0.20 $ & $ 2.69 \pm 0.08 $ & $ 1.96 \pm 0.11 $ \\
BLG176.6 & $  0.230 $ & $ -4.847 $ & $  2.31 \pm 0.17 $ & $ 2.90 \pm 0.07 $ & $ 1.84 \pm 0.09 $ & $ -1.66 \pm 0.21 $ & $ 2.61 \pm 0.07 $ & $ 2.15 \pm 0.10 $ \\
BLG176.7 & $  0.097 $ & $ -4.921 $ & $  2.14 \pm 0.27 $ & $ 2.82 \pm 0.09 $ & $ 2.07 \pm 0.11 $ & $ -1.20 \pm 0.21 $ & $ 2.95 \pm 0.07 $ & $ 1.93 \pm 0.11 $ \\
BLG176.8 & $ -0.037 $ & $ -4.994 $ & $  1.79 \pm 0.23 $ & $ 2.85 \pm 0.08 $ & $ 1.75 \pm 0.10 $ & $ -0.24 \pm 0.26 $ & $ 3.28 \pm 0.10 $ & $ 2.30 \pm 0.14 $ \\

%% file: paper.bbl
\begin{thebibliography}{36}
\expandafter\ifx\csname natexlab\endcsname\relax\def\natexlab#1{#1}\fi

\bibitem[{{Anderson} {et~al.}(2006){Anderson}, {Bedin}, {Piotto}, {Yadav}, \&
  {Bellini}}]{anderson06}
{Anderson}, J., {Bedin}, L.~R., {Piotto}, G., {Yadav}, R.~S., \& {Bellini}, A.
  2006, \aap, 454, 1029

\bibitem[{{Anderson} \& {King}(2000)}]{anderson00}
{Anderson}, J., \& {King}, I.~R. 2000, \pasp, 112, 1360

\bibitem[{{Babusiaux} {et~al.}(2010){Babusiaux}, {G{\'o}mez}, {Hill}, {Royer},
  {Zoccali}, {Arenou}, {Fux}, {Lecureur}, {Schultheis}, {Barbuy}, {Minniti}, \&
  {Ortolani}}]{babusiaux10}
{Babusiaux}, C., {G{\'o}mez}, A., {Hill}, V., {et~al.} 2010, \aap, 519, A77

\bibitem[{{Bellini} \& {Bedin}(2010)}]{bellini10}
{Bellini}, A., \& {Bedin}, L.~R. 2010, \aap, 517, A34

\bibitem[{{Bureau} {et~al.}(2006){Bureau}, {Aronica}, {Athanassoula},
  {Dettmar}, {Bosma}, \& {Freeman}}]{bureau06}
{Bureau}, M., {Aronica}, G., {Athanassoula}, E., {et~al.} 2006, \mnras, 370,
  753

\bibitem[{{Cao} {et~al.}(2013){Cao}, {Mao}, {Nataf}, {Rattenbury}, \&
  {Gould}}]{cao13}
{Cao}, L., {Mao}, S., {Nataf}, D., {Rattenbury}, N.~J., \& {Gould}, A. 2013,
  ArXiv e-prints 1303.6430

\bibitem[{{Clarkson} {et~al.}(2008){Clarkson}, {Sahu}, {Anderson}, {Smith},
  {Brown}, {Rich}, {Casertano}, {Bond}, {Livio}, {Minniti}, {Panagia},
  {Renzini}, {Valenti}, \& {Zoccali}}]{clarkson08}
{Clarkson}, W., {Sahu}, K., {Anderson}, J., {et~al.} 2008, \apj, 684, 1110

\bibitem[{{De Propris} {et~al.}(2011){De Propris}, {Rich}, {Kunder}, {Johnson},
  {Koch}, {Brough}, {Conselice}, {Gunawardhana}, {Palamara}, {Pimbblet}, \&
  {Wijesinghe}}]{depropis11}
{De Propris}, R., {Rich}, R.~M., {Kunder}, A., {et~al.} 2011, \apjl, 732, L36

\bibitem[{{Gallart}(1998)}]{gallart98}
{Gallart}, C. 1998, \apjl, 495, L43

\bibitem[{{Kuijken} \& {Rich}(2002)}]{kuijken02}
{Kuijken}, K., \& {Rich}, R.~M. 2002, \aj, 124, 2054

\bibitem[{{Kunder} {et~al.}(2012){Kunder}, {Koch}, {Rich}, {de Propris},
  {Howard}, {Stubbs}, {Johnson}, {Shen}, {Wang}, {Robin}, {Kormendy}, {Soto},
  {Frinchaboy}, {Reitzel}, {Zhao}, \& {Origlia}}]{kunder12}
{Kunder}, A., {Koch}, A., {Rich}, R.~M., {et~al.} 2012, \aj, 143, 57

\bibitem[{{Li} \& {Shen}(2012)}]{li12}
{Li}, Z.-Y., \& {Shen}, J. 2012, \apjl, 757, L7

\bibitem[{{Mao} \& {Paczy{\'n}ski}(2002)}]{mao02}
{Mao}, S., \& {Paczy{\'n}ski}, B. 2002, \mnras, 337, 895

\bibitem[{{McWilliam} \& {Zoccali}(2010)}]{mcwilliam10}
{McWilliam}, A., \& {Zoccali}, M. 2010, \apj, 724, 1491

\bibitem[{{Nataf} {et~al.}(2013){Nataf}, {Gould}, {Pinsonneault}, \&
  {Udalski}}]{nataf13}
{Nataf}, D.~M., {Gould}, A.~P., {Pinsonneault}, M.~H., \& {Udalski}, A. 2013,
  \apj, 766, 77

\bibitem[{{Nataf} {et~al.}(2010){Nataf}, {Udalski}, {Gould}, {Fouqu{\'e}}, \&
  {Stanek}}]{nataf10}
{Nataf}, D.~M., {Udalski}, A., {Gould}, A., {Fouqu{\'e}}, P., \& {Stanek},
  K.~Z. 2010, \apjl, 721, L28

\bibitem[{{Nataf} {et~al.}(2011){Nataf}, {Udalski}, {Gould}, \&
  {Pinsonneault}}]{nataf11}
{Nataf}, D.~M., {Udalski}, A., {Gould}, A., \& {Pinsonneault}, M.~H. 2011,
  \apj, 730, 118

\bibitem[{{Nataf} {et~al.}(2012){Nataf}, {Gould}, {Fouqu{\'e}}, {Gonzalez},
  {Johnson}, {Skowron}, {Udalski}, {Szyma{\'n}ski}, {Kubiak},
  {Pietrzy{\'n}ski}, {Soszy{\'n}ski}, {Ulaczyk}, {Wyrzykowski}, \&
  {Poleski}}]{nataf12tmp}
{Nataf}, D.~M., {Gould}, A., {Fouqu{\'e}}, P., {et~al.} 2012, ArXiv e-prints
  astro-ph/1208.1263

\bibitem[{{Ness} {et~al.}(2012){Ness}, {Freeman}, {Athanassoula},
  {Wylie-De-Boer}, {Bland-Hawthorn}, {Lewis}, {Yong}, {Asplund}, {Lane},
  {Kiss}, \& {Ibata}}]{ness12}
{Ness}, M., {Freeman}, K., {Athanassoula}, E., {et~al.} 2012, \apj, 756, 22

\bibitem[{{Ness} {et~al.}(2013){Ness}, {Freeman}, {Athanassoula},
  {Wylie-de-Boer}, {Bland-Hawthorn}, {Asplund}, {Lewis}, {Yong}, {Lane},
  {Kiss}, \& {Ibata}}]{ness13}
{Ness}, M., {Freeman}, K., {Athanassoula}, E., {et~al.} 2013, ArXiv e-prints
  1303.6656

\bibitem[{{P{\'a}l} \& {Bakos}(2006)}]{pal06}
{P{\'a}l}, A., \& {Bakos}, G.~{\'A}. 2006, \pasp, 118, 1474

\bibitem[{{Press} {et~al.}(1992){Press}, {Teukolsky}, {Vetterling}, \&
  {Flannery}}]{press92}
{Press}, W.~H., {Teukolsky}, S.~A., {Vetterling}, W.~T., \& {Flannery}, B.~P.
  1992, {Numerical recipes in C. The art of scientific computing} (UK:
  Cambridge University Press)

\bibitem[{{Rangwala} {et~al.}(2009){Rangwala}, {Williams}, \&
  {Stanek}}]{rangwala09}
{Rangwala}, N., {Williams}, T.~B., \& {Stanek}, K.~Z. 2009, \apj, 691, 1387

\bibitem[{{Reid} \& {Brunthaler}(2004)}]{reid04}
{Reid}, M.~J., \& {Brunthaler}, A. 2004, \apj, 616, 872

\bibitem[{{Robin} {et~al.}(2003){Robin}, {Reyl{\'e}}, {Derri{\`e}re}, \&
  {Picaud}}]{robin03}
{Robin}, A.~C., {Reyl{\'e}}, C., {Derri{\`e}re}, S., \& {Picaud}, S. 2003,
  \aap, 409, 523

\bibitem[{{Saito} {et~al.}(2011){Saito}, {Zoccali}, {McWilliam}, {Minniti},
  {Gonzalez}, \& {Hill}}]{saito11}
{Saito}, R.~K., {Zoccali}, M., {McWilliam}, A., {et~al.} 2011, \aj, 142, 76

\bibitem[{{Saito} {et~al.}(2012){Saito}, {Minniti}, {Dias}, {Hempel},
  {Rejkuba}, {Alonso-Garc{\'{\i}}a}, {Barbuy}, {Catelan}, {Emerson},
  {Gonzalez}, {Lucas}, \& {Zoccali}}]{saito12}
{Saito}, R.~K., {Minniti}, D., {Dias}, B., {et~al.} 2012, \aap, 544, A147

\bibitem[{{Sch{\"o}nrich}(2012)}]{schonrich12}
{Sch{\"o}nrich}, R. 2012, \mnras, 427, 274

\bibitem[{{Shen} {et~al.}(2010){Shen}, {Rich}, {Kormendy}, {Howard}, {De
  Propris}, \& {Kunder}}]{shen10}
{Shen}, J., {Rich}, R.~M., {Kormendy}, J., {et~al.} 2010, \apjl, 720, L72

\bibitem[{{Spaenhauer} {et~al.}(1992){Spaenhauer}, {Jones}, \&
  {Whitford}}]{spaenhauer92}
{Spaenhauer}, A., {Jones}, B.~F., \& {Whitford}, A.~E. 1992, \aj, 103, 297

\bibitem[{{Sumi} {et~al.}(2004){Sumi}, {Wu}, {Udalski}, {Szyma{\'n}ski},
  {Kubiak}, {Pietrzy{\'n}ski}, {Soszy{\'n}ski}, {Wo{\'z}niak},
  {{\.Z}ebru{\'n}}, {Szewczyk}, \& {Wyrzykowski}}]{sumi04}
{Sumi}, T., {Wu}, X., {Udalski}, A., {et~al.} 2004, \mnras, 348, 1439

\bibitem[{{Szyma{\'n}ski} {et~al.}(2011){Szyma{\'n}ski}, {Udalski},
  {Soszy{\'n}ski}, {Kubiak}, {Pietrzy{\'n}ski}, {Poleski}, {Wyrzykowski}, \&
  {Ulaczyk}}]{szymanski11}
{Szyma{\'n}ski}, M.~K., {Udalski}, A., {Soszy{\'n}ski}, I., {et~al.} 2011,
  \actaa, 61, 83

\bibitem[{{Udalski}(2003)}]{udalski03}
{Udalski}, A. 2003, \actaa, 53, 291

\bibitem[{{Uttenthaler} {et~al.}(2012){Uttenthaler}, {Schultheis}, {Nataf},
  {Robin}, {Lebzelter}, \& {Chen}}]{uttenthaler12}
{Uttenthaler}, S., {Schultheis}, M., {Nataf}, D.~M., {et~al.} 2012, \aap, 546,
  A57

\bibitem[{{Vieira} {et~al.}(2007){Vieira}, {Casetti-Dinescu}, {M{\'e}ndez},
  {Rich}, {Girard}, {Korchagin}, {van Altena}, {Majewski}, \& {van den
  Bergh}}]{vieira07}
{Vieira}, K., {Casetti-Dinescu}, D.~I., {M{\'e}ndez}, R.~A., {et~al.} 2007,
  \aj, 134, 1432

\bibitem[{{Whitmore} \& {Bell}(1988)}]{whitmore88}
{Whitmore}, B.~C., \& {Bell}, M. 1988, \apj, 324, 741

\end{thebibliography}
